\documentclass[aps,prd,preprint,superscriptaddress,tightenlines,%
showpacs,nofootinbib]{revtex4}
\newcommand{\PRE}[1]{{#1}}   

\usepackage{epsfig}

\newcommand{\postscript}[2]{\setlength{\epsfxsize}{#2\hsize}
   \centerline{\epsfbox{#1}}}

\newcommand{\md}{M_D}
\newcommand{\mbh}{M_{\text{BH}}}

\newcommand{\xmin}{x_{\text{min}}}

\newcommand{\gev}{\text{GeV}}
\newcommand{\tev}{\text{TeV}}

\newcommand{\etal}{{\em et al.}}

\newcommand{\eqref}[1]{Eq.~(\ref{#1})}
\def\mah{M_{\rm A.H.}}

\begin{document}

\preprint{
\hfil
\begin{minipage}[t]{3in}
\begin{flushright}
\vspace*{.4in}
NUB--3239--Th--03\\
UCI--TR--2003--31\\
UK/03--10\\
hep-ph/0307228
\end{flushright}
\end{minipage}
}

\title{ \PRE{\vspace*{1.5in}} 
Updated Limits on TeV--Scale Gravity from 
Absence of Neutrino Cosmic Ray Showers 
Mediated by Black Holes\PRE{\vspace*{0.3in}} }

\author{Luis A.~Anchordoqui}
\affiliation{Department of Physics,
Northeastern University, Boston, MA 02115
\PRE{\vspace*{.1in}}
}

\author{Jonathan L.~Feng}
\affiliation{Department of Physics and Astronomy,
University of California, Irvine, CA 92697
\PRE{\vspace*{.1in}}
}

\author{Haim Goldberg}
\affiliation{Department of Physics,
Northeastern University, Boston, MA 02115
\PRE{\vspace*{.1in}}
}

\affiliation{Center for Theoretical Physics,
Massachusetts Institute of Technology, Cambridge, MA 02139
\PRE{\vspace*{.1in}}
}

\author{Alfred D.~Shapere%
\PRE{\vspace*{.3in}}
}
\affiliation{Department of Physics,
University of Kentucky, Lexington, KY 40506
\PRE{\vspace*{.5in}}
}


\begin{abstract}
\PRE{\vspace*{.1in}} We revise existing limits on the $D$-dimensional
Planck scale $\md$ from the nonobservation of microscopic black holes
produced by high energy cosmic neutrinos in scenarios with $D=4{+}n$
large extra dimensions. Previous studies have neglected the energy
radiated in gravitational waves by the multipole moments of the
incoming shock waves. We include the effects of energy loss, as well
as form factors for black hole production and recent null results from
cosmic ray detectors.  For $n \geq 5$, we obtain $\md > 1.0 -
1.4~\tev$. These bounds are among the most stringent and conservative
to date.
\end{abstract}


\maketitle

\section{Introduction}

Forthcoming colliders~\cite{Giddings:2001bu}, cosmic ray
observatories~\cite{Feng:2001ib,Anchordoqui:2001ei,Emparan:2001kf},
neutrino telescopes~\cite{Kowalski:2002gb}, and space-based
experiments~\cite{Dutta:2002ca} will be able to observe black holes
(BHs) if the fundamental scale of gravity is sufficiently close to 1
TeV~\cite{Antoniadis:1998ig}.  Observations of highly characteristic
BH events at any of these facilities could conceivably provide the
first evidence for the existence of extra dimensions and make possible
the direct study of strong quantum gravity effects and strings.  On
the other hand, the lack of such events in any experiments to date
leads to lower limits on the scale of higher-dimensional
gravity~\cite{Anchordoqui:2001cg}.

To make useful predictions about higher-dimensional gravity based on
observations of such events, or their absence, a quantitative
understanding of the process of BH production in high-energy
collisions is required.  An intuitive picture of this process is
provided by a simple model known as Thorne's hoop
conjecture~\cite{Thorne:ji}, according to which a BH forms in a
two-particle collision when and only when the impact parameter is
smaller than the radius $r_s$ of a Schwarzschild BH of mass equal to
the total center-of-mass energy $E_{\rm CM}$.  The hoop conjecture
thus predicts a total cross section for BH production equal to the
area subtended by a ``hoop'' of radius $r_s$:
\begin{equation}
\sigma_{BH}^{\text{hoop}} = \pi r_s^2(E_{\rm CM}) \ .
\label{hoopsigma}
\end{equation}  
Up to now, all studies of BH production in TeV-scale gravity have been based
on this rather heuristic cross section, and have thus been subject to
substantial theoretical uncertainties.

Relatively recently, significant progress has been made in determining
the cross section for BH production.  Early analytic calculations in
four dimensions~\cite{Penrose,D'Eath:hb} for head-on collisions
illustrated the process of horizon formation and found that the mass
of the final BH was about 84\% of the initial center-of-mass
energy. These calculations were extended to nonzero impact parameter
by Eardley and Giddings~\cite{Eardley:2002re}, who analytically
derived a lower bound on the total cross section of approximately 65\%
of Eq.~(\ref{hoopsigma}).  Relatively recently, a calculation of the
cross section in higher dimensions was performed by Yoshino and Nambu
using numerical techniques~\cite{Yoshino:2002br}.  In addition, these
authors observed significant reductions in the mass of the final-state
black hole as a function both of impact parameter and dimension.

If TeV-scale gravity is realized in nature, then the first
observational evidence for it will likely come from BH-mediated
neutrino cosmic ray showers.  Ultra-high energy cosmic rays hit the
Earth with collision center-of-mass energies ranging up to roughly
$10^5~\gev$. QCD cross sections dominate over the BH production cross
section by a factor of roughly $10^9$. Thus, black holes produced by
hadronic cosmic rays are effectively unobservable. This is not the
case for incoming neutrinos, whose cross section for producing black
holes can be orders of magnitude larger than SM cross sections, but
much less than hadronic~\cite{Feng:2001ib}. As a consequence,
neutrinos interact with roughly equal probability at any point in the
atmosphere, and the light descendants of the black hole may initiate
quasi-horizontal showers in the volume of air immediately above the
detector.  Because of these considerations the atmosphere provides a
buffer against contamination by mismeasured hadrons, allowing a good
characterization of BH-induced showers when the BH entropy $S \gg
1$~\cite{Anchordoqui:2001ei}.  Additionally, neutrinos that traverse
the atmosphere unscathed may produce black holes through interactions
in the ice or water~\cite{Kowalski:2002gb}.  Because the BH production
cross section is suppressed by a power of the fundamental Planck scale
$\md$ (approaching $\md^2$ for large numbers of extra dimensions), the
absence of neutrino showers mediated by black holes implies lower
bounds on $\md$.

In this article, we bring up to date existing
limits~\cite{Anchordoqui:2001cg,Anchordoqui:2002vb} on $\md$ from the
nonobservation of BHs at cosmic neutrino detection
experiments. Besides incorporating the cross section and energy loss
results of Yoshino and Nambu, we also make use of updated parton
distribution functions and recently available cosmic ray data.

\section{Energy Loss in Black Hole Creation}

Previous calculations of the cross section for producing a BH have
neglected energy loss in the creation of a BH, assuming that the mass
of the created black hole $M_{\rm BH}$ was identical to the incoming
parton center-of-mass energy $\sqrt{\hat{s}}$.  However, recent
work~\cite{Yoshino:2002br} has shown that the energy lost to
gravitational radiation is not negligible, and in fact is large for
larger $n$ and for large impact parameters.  The trapped mass (called
$M_{\rm A.H.}$ in Ref.~\cite{Yoshino:2002br}, and which we continue to
call $M_{\rm BH}$~\cite{Hawking}), is given by
\begin{equation}
M_{\rm BH}(z) = y(z)\sqrt{\hat s} \ ,
\end{equation}
where the inelasticity $y$ is a function of $z\equiv b/b_{\rm max}$.
Here $b$ is the impact parameter and
\begin{equation}
b_{\rm max}=\sqrt{F(n)} \ r_s(\sqrt{\hat s}, n, \md)
\end{equation}
is the maximum impact parameter for collapse, where
\begin{equation}
\label{schwarz}
r_s(\sqrt{\hat{s}}, n, \md) =
\frac{1}{\md}
\left[ \frac{\sqrt{\hat s}}{\md} \right]^{\frac{1}{1+n}}
\left[ \frac{2^n \pi^{(n-3)/2}\Gamma({n+3\over 2})}{n+2}
\right]^{\frac{1}{1+n}}
\end{equation}
is the radius of a Schwarzschild BH in $(4{+}n)$
dimensions~\cite{Myers:un}, and $F(n)$ is the form factor explicitly
given in Ref.~\cite{Yoshino:2002br}.

This complicates the parton model calculation, since the production of
a BH of mass $M_{\rm BH}$ requires that $\hat s$ be $M_{\rm
BH}^2/y^2(z)$, thus requiring the lower cutoff on parton momentum
fraction to be a function of impact parameter~\cite{note1}.  In what
follows we take the $\nu N$ cross section as an impact
parameter-weighted average over parton cross sections, with the lower
parton fractional momentum cutoff determined by the requirement
$M_{\rm BH}^{\text{min}}=x_{\rm min}M_D$~\cite{note3/2}. This gives a
lower bound $x^2_{\rm min} M_D^2/[y^2(z)\,s]$ on the parton momentum
fraction $x$. With this in mind, the $\nu N \to {\rm BH}$ cross
section is
\begin{equation}
\sigma(E_\nu,\xmin,n,M_D) \equiv \int_0^1 2 z \,dz 
\int_{\frac{(\xmin \md)^2}{y^2s}}^1 dx \, F(n) \,  
\pi r_s^2(\sqrt{\hat{s}}, n, \md) \,\sum_i f_i(x,Q) \ ,
\label{sigma}
\end{equation}  
where $\xmin$ is determined by the requirement that the BH have at
least an approximate semi-classical description, $\hat{s} = 2 x m_N
E_\nu$, $i$ labels parton species, and the $f_i(x,Q)$ are parton
distribution functions (pdfs)~\cite{Anchordoqui:2001cg}. 

The choice of the momentum transfer $Q$ is governed by considering the
time or distance scale probed by the interaction. Roughly speaking,
the formation of a well-defined horizon occurs when the colliding
particles are at a distance $\sim r_s$ apart. This has led to the
advocacy of the choice $Q \simeq r_s^{-1}$~\cite{Emparan:2001kf},
which has the advantage of a sensible limit at very high
energies. However, the dual resonance picture of string theory would
suggest a choice $Q \sim \sqrt{\hat s}$. Fortunately, as noted in
Refs.~\cite{Anchordoqui:2001cg,Anchordoqui:2002vb}, the BH production
cross section is largely insensitive to the details of the choice of
$Q$. In what follows we use the CTEQ6M pdfs~\cite{Pumplin:2002vw} with
$Q = {\rm min} \{r_s^{-1}, 10~{\rm TeV}\}$.

\begin{figure}
\postscript{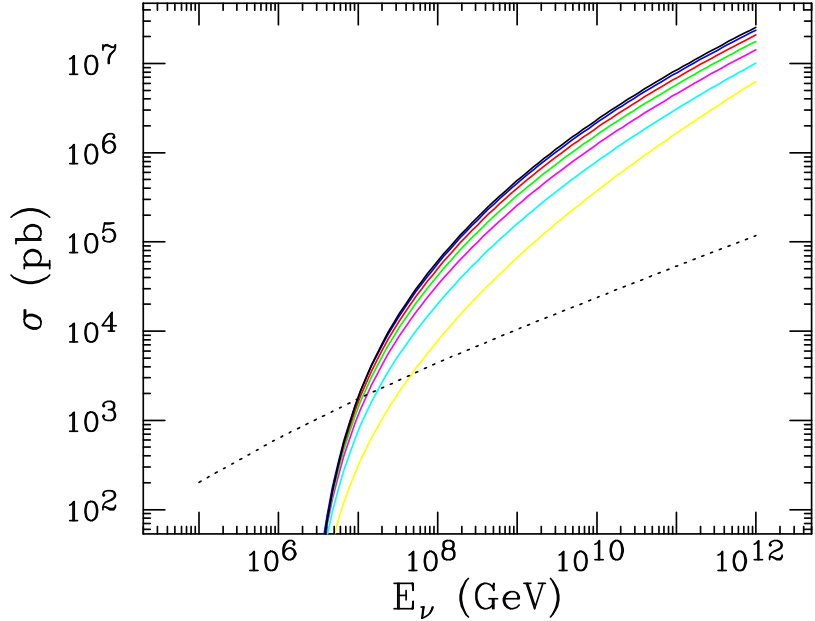}{0.65}
\caption{Cross sections $\sigma(\nu N \to {\rm BH})$ for $n=1,\ldots,
7$ from below, assuming $\md = 1~\tev$ and $\xmin=1$.  Energy loss has
been included according to \eqref{sigma}.  The SM cross section
$\sigma(\nu N \to \ell X)$ is indicated by the dotted line.}
\label{sigmaf}
\end{figure}

In Fig.~\ref{sigmaf} we show the BH production cross section for $n =
1, \dots, 7$ extra dimensions with energy loss incorporated as given
in \eqref{sigma}.  The rapid rise in cross section is pushed to higher
$E_\nu$ than in the case with energy loss neglected.  However, the
cross sections are still well above the SM cross section at $E_{\nu}
\sim 10^8~\gev$ and above, where, as we will see, the cosmogenic
neutrino flux is large.

\section{Cosmic Neutrino Detectors}

Energy loss also impacts event rates at cosmic neutrino detectors, not
only because the cross section is modified, but also because the
apertures of cosmic neutrino detectors are functions of shower energy.
Let $N_A$ be Avogadro's number, $A(y E_\nu)$ the neutrino aperture of
a given experiment for shower energy $y E_{\nu}$, and $T$ be the
experiment's running time.  The number of neutrino showers mediated by
BHs is then
\begin{eqnarray}
{\cal N} (x_{\text{min}}, n, M_D) 
&=& N_A T \int dE_{\nu} \int_0^1 2 z\, dz
\int_{\frac{(x_{\text{min}} M_D)^2}{y^2 s}}^1 dx \,
\frac{d\Phi}{dE_{\nu}} \, A(y E_{\nu}) \nonumber \\
&& \times F(n) \, \pi r_s^2(\sqrt{\hat{s}}, n, M_D) \sum_i f_i(x,Q) \ ,
\label{bhevents}
\end{eqnarray}
where $d\Phi/dE_\nu$ is the diffuse flux of cosmic neutrinos hitting
the Earth.

There are several techniques employed in detecting neutrino
showers~\cite{Sigl:2001th}. The most commonly used method involves
giant arrays of particle counters that sample the lateral and temporal
density profiles of the muon and electromagnetic components of the
shower front. Another well-established method involves measurement of
the air shower evolution --- its growth and subsequent attenuation ---
as it develops by sensing the fluorescence light produced via
interactions of the charged particles in the atmosphere. A third
method exploits naturally occurring large volume \v{C}erenkov
radiators such as deep water or ice.  Especially useful at relevant
center-of-mass energies for BH production is emission of \v{C}erenkov
radiation at radio frequencies.  For fluorescence data, a direct
measurement of the depth of shower maximum $X_{\rm max}$ and the shape
of the longitudinal profile provide sensitive diagnostics in
discrminating between neutrino and hadron showers. In the case of
surface arrays, the composition information is extracted from a number
of shower characteristics which reflect the depth of shower maximum
and the ratio of muon to electromagnetic content of the shower.

The AGASA Collaboration~\cite{Yoshida} reports no significant
enhancement of deeply-developing shower rates given the detector's
resolution. Specifically, there is only 1 event observed, consistent
with the expected background of 1.72 from hadronic cosmic rays. For
details, see Ref.~\cite{Anchordoqui:2001cg}.

The Fly's Eye detector ceased operation in July 1992 after a life of
11 years. It was designed to collect the atmospheric nitrogen
fluorescence light produced by air shower particles on moonless nights
without cloud cover, achieving an overall duty cycle of $\approx
10\%$. The experiment recorded more than 5000 events, but no unusual
deeply developing showers have been
found~\cite{Baltrusaitis:mt}. 

Recently, data from an upscaled version of the Fly's Eye experimet
have become available~\cite{Archbold:rp}.  The effective aperture of
the High Resolution Fly's Eye detector is on average about 6 times the
Fly's Eye aperture, with a threshold around $10^8$~GeV. The instrument
includes two sites (HiRes I and II) located 12.6~km apart. Each site
consists of a large number (22 at HiRes I and 42 at HiRes II) of
telescope units pointing at different parts of the sky. Between
November 1999 and September 2001, 1198 events were recorded with at
least one reconstructed energy greater than
$10^{8.7}~\gev$~\cite{Archbold:rp}. Because of bad weather conditions,
272 events were discarded from the sample. None of the 723 events that
survived all of the cuts required for stereo-mode triggering has
$X_{\rm max} > 1200$~g/cm$^2$. Additionally, there are no events
detected in monocular mode with $X_{\rm max} > 1500$~g/cm$^2$. In the
spirit of Ref.~\cite{Yoshida:1996ie}, we parametrize the HiRes
aperture for deeply ($X_{\rm max} > 1500$~g/cm$^2$) developing showers
by
\begin{equation}
A (E_\nu) = 1.8 \, \left\{\left[2.7 
+ \log \left(\frac{E_\nu}{10^{10}~{\rm GeV}}\right)\right]^2 
- 0.5\right\} \, {\rm km}^{3} {\rm we}\,{\rm sr}  \ .
\end{equation}
We note that there is an additional small contribution to the HiRes
exposure in the energy range $10^8 - 10^9$~GeV from data collected
during 2878 hours livetime~\cite{Abu-Zayyad:1999xa}.

The Radio Ice \v{C}erenkov Experiment (RICE) is designed to detect the
radio frequency \v{C}erenkov radiation produced by neutrino-induced
showers in ice~\cite{Frichter:1995cn}.  Specifically, the
electromagnetic channel of the shower produces a radio pulse with a
duration of a few nanoseconds and with power concentrated around the
\v{C}erenkov angle. Several radio antennae positioned in the ice allow
for reconstruction of the interaction vertex.  For primary energies
above $10^9~\gev$, the Landau Pomeranchuk-Migdal (LPM)
effect~\cite{Landau:um} leads to a significant suppression of the
Bethe-Heitler cross sections for the pair production and
Bremsstrahlung processes in dense materials, and thus dramatically
changes the character of the development of electromagnetic showers.

Almost instantaneously after its formation, the TeV-scale BH
decays~\cite{Hawking:1975sw}, predominantly through radiation of
standard model (SM) particles~\cite{Emparan:2000rs}.  About 75\% of
the BH energy is carried off by quarks and gluons and rougly a third
of this energy goes into the electromagnetic channel via $\pi^0$
decay.  Only about 5\% of particles directly emitted from the BH
($\nu$'s, $\tau$'s, $\mu$'s) do not partake in the shower. The rest of
the energy eventually devolves into secondary electromagnetic cascades
with particle energies below that for which the LPM effect is
important~\cite{note2}.  As a conservative estimate we model the
aperture for neutrino showers mediated by BHs using the hadronic
effective volume reported by the RICE
Collaboration~\cite{Kravchenko:2002mm} with average inelasticity
$\langle y \rangle =0.8$. This estimate is supported by the fact that
BH-induced showers mimic SM neutral current events, characterized by
hadronic dominated showers with no leading charged lepton. A more
rigorous analysis of the BH acceptance at the RICE facility is
underway~\cite{McKay}.

The RICE detector comprises 16 dipole radio receivers installed in the
holes drilled for the AMANDA experiment in the Antaractic ice.  Four
transmitter antennae are also deployed in the ice for calibration
purposes.  The trigger requires a 4-fold coincidence within a
1.2~$\mu$s time window, and various cuts are applied to reject thermal
and anthropogenic backgrounds.  For example, shower vertices are
reconstructed using a $\chi^2$ fit to the signal arrival times, and
the resulting fits are required to be of sufficient quality and to
indicate vertices at least 50~m below the ice's surface.  During a 1
month run in August of 2000 with a livetime of 333.3 hours, a total of
22 events passed all the automated cuts.  These events were then
scanned for quality and the RICE Collaboration concluded that there
are no events consistent with neutrino
sources~\cite{Kravchenko:2002mm}.

\begin{figure}
\postscript{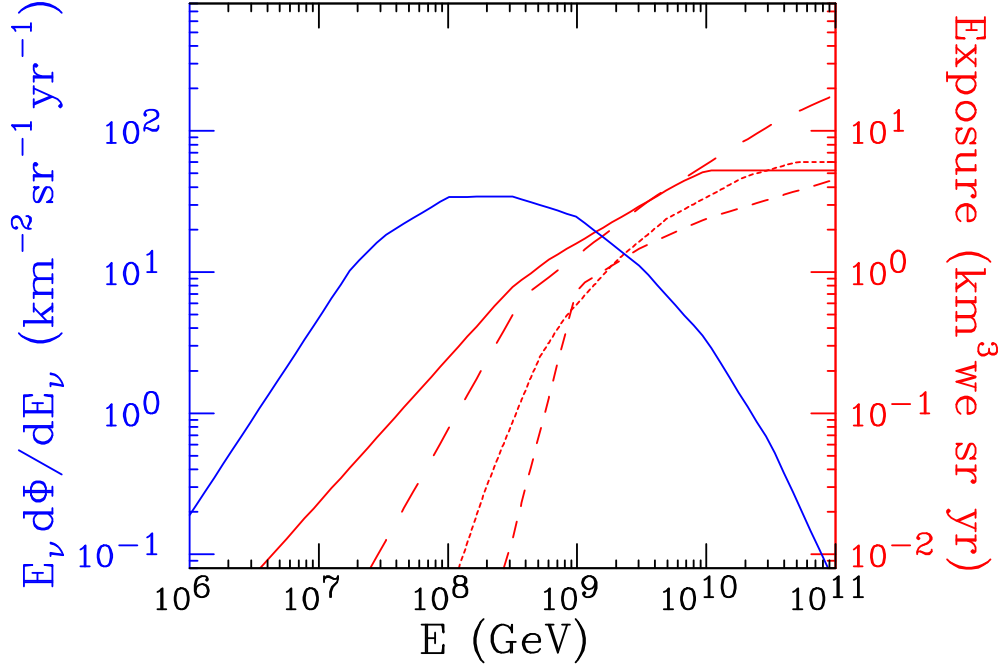}{0.80}
\caption{The monotonically rising curves are the exposures as
functions of shower energy for AGASA (solid), Fly's Eye (dotted),
HiRes (short dash), and RICE (long dash).  The remaining solid curve,
with a peak around $10^{8.5}$~GeV, is the cosmogenic neutrino flux.}
\label{fexposure}
\end{figure}

The relative exposures for the different experiments are given in
Fig.~\ref{fexposure}. For details on the apertures of AGASA and Fly's
Eye, the reader is referred to our previous
paper~\cite{Anchordoqui:2002vb}.  All in all, there is only 1 event
observed with an expected background of 1.72 from hadronic cosmic
rays, leading to a 95\% CL limit of 3.5 BH
events~\cite{Feldman:1997qc}.

To derive the bounds on $M_D$, we use the ``guaranteed'' flux of
cosmogenic neutrinos arising from the decay of $\pi^\pm$ produced in
collisions of ultra-high energy protons with the cosmic microwave
background. As in our previous analyses, we conservatively adopt the
estimates of Protheroe and Johnson~\cite{Protheroe:1996ft} with
nucleon source spectrum scaling as $d\Phi_N / dE \propto E^{-2}$ and
extending up to the cutoff energy $10^{12.5}~\gev$. The total
ultra-high energy cosmogenic neutrino flux is also shown in
Fig.~\ref{fexposure}.

\section{Bounds}

\begin{figure}
\postscript{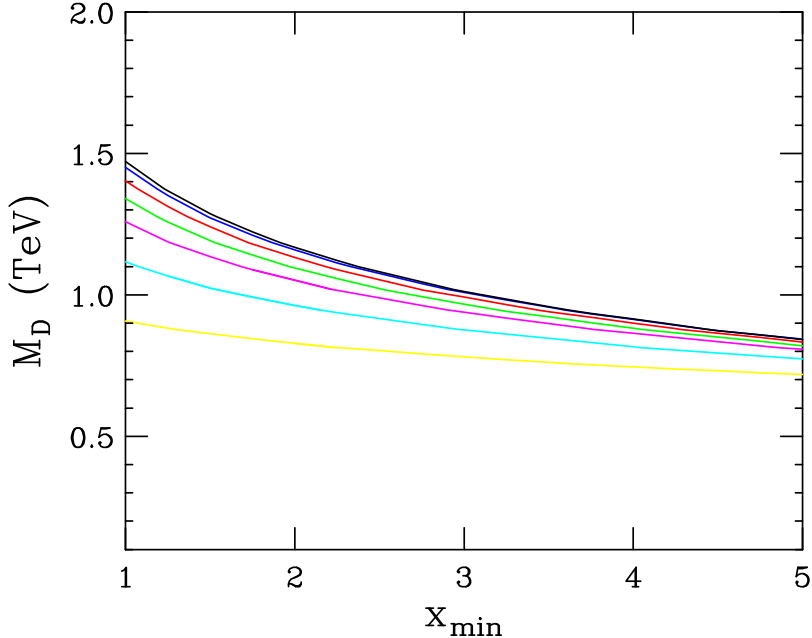}{0.65}
\caption{95\% CL lower limits on the fundamental Planck scale as a
function of $\xmin$ for $n = 1, \ldots, 7$ extra dimensions (from
below).}
\label{limits}
\end{figure}

In Fig.~\ref{limits} we show 95\% CL lower bounds on $\md$ as derived
from Eq.~(\ref{bhevents}) using the exposures and the cosmogenic
neutrino flux given in Fig.~\ref{fexposure}, requiring ${\cal N} <
3.5$ events to be observed in cosmic neutrino data
samples~\cite{Giddings:2000ay}. The BH entropy is a measure of the
validity of the semi-classical approximation.  For $\xmin \agt 3$ and
$n\geq 5$, the entropy
\begin{equation}
S = \frac{4 \pi\, \mbh\,r_s(\mbh)}{n+2} \gg 10  \ ,
\end{equation}
yielding small thermal fluctuations in the emission
process~\cite{Preskill:1991tb}. Hence, for $\xmin \agt 3$ and $n\geq
5$, strong quantum gravity effects may be safely neglected. Moreover,
gravitational effects due to brane back-reaction are expected to be
insignificant for $\mbh$ well beyond the brane tension, which is
presumably of the order of $\md$.  The uncertainty illustrated in
Fig.~\ref{limits} associated with $\xmin$ only concerns BH production
and is highly insensitive to decay
characteristics~\cite{Anchordoqui:2002cp}.  This is because the signal
in neutrino detection experiments relies only on the existence of
visible decay products.  Whatever happens around $\xmin \approx 1$, it
seems quite reasonable to expect that BHs or their Planckian
progenitors will cascade decay on the brane.

String theory provides a more complete picture of the decay for $\mbh$
close to $\md$, which may further justify setting $\xmin =1$.  (Such
arguments do not address the issue of brane back-reaction, however.)
In string theory, the ultimate fate of the black hole is determined by
the string/BH correspondence principle~\cite{Susskind:ws}: when the
Schwarzschild radius of the black hole shrinks to the fundamental
string length $\ell_s \gg \ell_D$, where $\ell_D$ is the fundamental
$(4+n)$-dimensional Planck length, an adiabatic transition occurs to a
massive superstring mode. Subsequent energy loss continues as thermal
radiation at the unchanging Hagedorn
temperature~\cite{Amati:1999fv}. The continuity of the cross section
at the correspondence point, parametrically in both the energy and the
string coupling, provides independent support for this
picture~\cite{Dimopoulos:2001qe}. Thus, the cross sections given in
Fig.~\ref{sigmaf} can be thought of as lower bounds on $\sigma$ as
$\mbh$ approaches $\md$~\cite{Cheung:2002aq}.

\section{Summary}

Incorporation of the results of Ref.~\cite{Yoshino:2002br} has
eliminated many of the sources of uncertainty enumerated in
Ref.~\cite{Anchordoqui:2001cg} and recapitulated in
Ref.~\cite{Ahn:2003qn}.  In Ref.~\cite{Anchordoqui:2001cg} we
identified two sources of uncertainty that could reduce the total
cross section: the reduction of the mass of the final-state BH
relative to the initial center-of-mass energy, and expectations for a
reduced cross section at nonzero center-of-mass angular
momentum.\footnote{In fact, a slight modification of our estimate for
the modification of the cross section due to angular momentum effects
has been used to give a surprisingly accurate postdiction of the
higher-dimensional cross section for BH production~\cite{Ida:2002ez},
providing further evidence for the correctness of the Yoshino and
Nambu calculations.}  On the other hand, we pointed out that the
classical photon capture cross section and nonrelativistic estimates
suggest a possible enhancement to the na\"{\i}ve geometric cross
section $\pi r_s^2$ by a factor of 2 or more. (The claim
of~\cite{Ahn:2003qn} that this upside uncertainty casts doubt on the
program of setting limits on $M_D$ from nonobservation of BHs is
mistaken.) Thus we concluded that, in the absence of a better
quantitative understanding of the process of BH formation, the
na\"{\i}ve geometric cross section provided a reasonable estimate.

With such calculations now in hand~\cite{Yoshino:2002br} we have
repeated our analysis and eliminated much of the uncertainty contained
in our previous limits on $M_D$, as well as incorporating updated
exposures from the HiRes and RICE facilities and updated
pdfs. (Incidentally, using the new pdfs contributed a net difference
of about 2\% to our results, confirming our previous
claim~\cite{Anchordoqui:2001cg}\ that there is very little sensitivity
to different choices of pdfs.  
Furthermore, the bulk of the sensitivity is for pdfs at $x>
m_{\text{BH}}^2/(2m_N E_\nu) \sim 10^{-2}$ and large $Q$, where the
pdfs are expected to be quite accurate.)  
In the course of our analysis we observed a
competition of effects leading to corrections to our previous
estimates: enhancement of the geometric cross section by form factors
of up to 1.9 and enhancement of apertures from new cosmic ray data,
but a simultaneous reduction in the rate of production of BHs of mass
greater than $x_{\rm min}M_D$, after taking energy losses into
account.  It turns out that the latter effect dominates and leads to a
slight weakening of our limits on $M_D$.  At the same time, our limits
are now on a much firmer theoretical footing, and maximally
conservative in all respects.

\begin{figure}[tbp]
\postscript{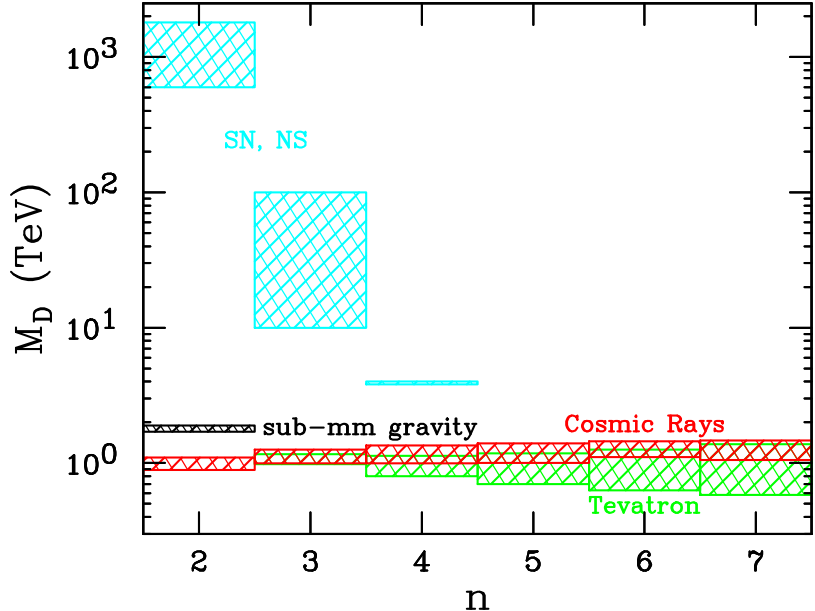}{0.65}
\caption{Bounds on the fundamental Planck scale $\md$ from tests of
Newton's law on sub-millimeter scales, bounds on supernova cooling and
neutron star heating, dielectron and diphoton production at the
Tevatron, and nonobservation of BH production by cosmic neutrinos.
The uncertainty in the Tevatron bounds corresponds to the range of
brane softening parameter $\in\!(\md/2, \md)$; for details see
Ref.~\cite{Anchordoqui:2001cg}. The range in the cosmic ray bounds is
for $\xmin= 1 - 3$.}
\label{summary}
\end{figure}

In Fig.~\ref{summary} we compare the bounds derived in this article
with existing limits on the fundamental scale of large extra
dimensions.  Tests of the gravitational inverse-square law at length
scales well below 1~mm show no evidence for short range Yukawa
interactions.  For $n=2$, this negative result can be translated into
a 95\% CL upper limit of 150~$\mu$m on the compactification radius of
flat extra dimensions, or equivalently to a unification mass scale
$\md > 1.8$~TeV~\cite{Hoyle:2000cv}. In such toroidal
compactifications the accessibility of towers of Kaluza-Klein
gravitons may drastically affect the phenomenology of supenovae and
neutron stars. For $n \leq 3$, anomalous cooling of supernovae due to
bulk graviton emission and neutron star heating by decay of
gravitationally trapped Kaluza-Klein modes provide limits on $\md$
that greatly exceed 1~TeV~\cite{Cullen:1999hc}. 

For $n\geq 4$ the sensitivity of table-top experiments and
astrophysical observations to TeV-scale gravity is largely reduced:
already for $n=4$ ($n=5$) supernova cooling yields $M_D > 4.0$~TeV
($\md > 0.8$~TeV)~\cite{Hewett:2002hv}.  For $n \geq 5$, the best
existing limits on TeV-scale gravity are from the absence of
trans-Planckian signatures (BH/stringball production) in neutrino
detection experiments discussed here, and from searches for
sub-Planckian signatures (graviton emission and virtual graviton
exchanges) at the Tevatron~\cite{Abbott:2000zb} and
LEP~\cite{Acciarri:1999jy}.  For $n \geq 5$ we have derived
conservative bounds incorporating the lower limits on the the mass
trapped in the gravitational collapse.  The resulting bounds, $\md >
1.0 - 1.4$~TeV for $\xmin =1-3$, are competitive with those obtained
in colliders and among the most stringent to date.

\begin{acknowledgments}
HG thanks the Aspen Center for Physics for hospitality during the
performance of some of this research.  The work of LAA and HG has been
partly supported by the US National Science Foundation (NSF), under
grants No.\ PHY--0140407 and No.\ PHY--0073034, respectively. The work
of JLF is supported in part by NSF CAREER grant No.\ PHY--0239817. The
work of ADS is supported in part by Department of Energy grant No.\
DE--FG01--00ER45832 and NSF grants PHY--0071312 and PHY--0245214.
\end{acknowledgments}



\begin{thebibliography}{99}
\bibitem{note1} The validity of an impact parameter description in the
semi-classical calculation can be ascertained by calculating the
maximum angular momentum $J_{\rm max}=(\sqrt{\hat s}/2)b_{\rm max}$ in
a given collision: for $M_{\rm BH}\ge M_D, $ one finds $J_{\rm max}\ge
7 \gg 1$.
\bibitem{note3/2} Because of the complexity of the final state, we
assume that amplitude intereference effects can be ignored. This
permits the cross section to be written as an integration over impact
parameter.  One of us (H.G.) would like to thank Scott Thomas for a
discussion on this point.
\bibitem{note2} At relevant energies ($10^9 - 10^{11}$~GeV) the mean
free path for $\pi$-ice interaction is roughly 120~cm, whereas the
charged pion decay length is about 780 $\gamma$~cm, where $\gamma$ is
the Lorentz factor. This implies that there is a negligible
probability for pion decay and so the shower cools to energies where
the LPM effect is small.
\end{thebibliography}
\end{document}